\documentclass[prd,showpacs,nofootinbib,tightenlines,%
preprintnumbers,superscriptaddress,%
twocolumn
]{revtex4}
\usepackage{epsfig}
\usepackage{amssymb}
\usepackage{amsfonts}
\usepackage{subfigure}
\usepackage{float}
\usepackage{latexsym,hyperref}
\graphicspath{{./Figures/}} 
\usepackage{graphics}
\usepackage{amssymb,amsmath,amsthm,graphicx}

\begin{document}
\preprint{HU-EP-06/30}

%-----------------------------------------------------------------------
\title{Modified $SO(3)$ Lattice Gauge Theory at $T\ne0$ with Parallel 
Tempering: \\ Monopole and Vortex Condensation}
%-----------------------------------------------------------------------

\author{G.~Burgio\footnote{Address from September 1$^{\rm st}$, 2006: Institut
f\"ur Theoretische Physik, Auf der Morgenstelle 14, D-72076 T\"ubingen, 
Germany}, M.~Fuhrmann, W.~Kerler, and M.~M\"uller-Preussker}
\affiliation{Humboldt-Universit\"at zu Berlin, Institut f\"ur Physik, 
Newtonstr.~15,
D-12489 Berlin, Germany}

\date{December 29, 2006}

\begin{abstract}
%---------------
The deconfinement transition is studied close to
the continuum limit of $SO(3)$ lattice gauge theory. High barriers for 
tunnelling among different twist sectors causing loss of ergodicity 
for local update algorithms are circumvented by means of parallel tempering. 
We compute monopole and center vortex free energies both within the confining 
phase and through the deconfinement transition. We discuss in detail the 
general problem of defining order parameters for adjoint actions. 
\end{abstract}


\maketitle 

\section{Introduction}
%---------------------
Understanding confinement in $SU(N)$ Yang-Mills theories remains one of the
major challenges of contemporary particle physics. Lattice simulations
have offered unique insight into the non-perturbative regularization of
pure gauge actions transforming under the fundamental representation of $SU(N)$
\cite{McLerran:1981pk,Kuti:1981gh}, equivalent to the quenched limit of 
full QCD: at non-zero temperature they have been shown to possess a phase 
transition linked to the spontaneous breaking of center symmetry 
\cite{Polyakov:1978vu,Susskind:1979up}.
For $N=2$ it is of second order, therefore lying in the universality 
class of the 3-d Ising model. However the question whether and in what sense 
this also holds for discretizations transforming under the natural continuum 
{\it pure} Yang-Mills gauge symmetry group $SU(N)/\mathbb{Z}_N$, for $N=2$ 
equivalent to $SO(3)$, still needs to be appropriately answered 
\cite{Smilga:1993vb}. According to universality \cite{Svetitsky:1982gs}, 
i.e. expecting the different formulations to be equivalent in the continuum 
limit, they should lead to the same non-perturbative physics. A discretization 
which does not break the $SU(N)/\mathbb{Z}_N$ invariance has moreover the 
appeal to preserve the topological properties related to 
$\pi_1(SU(N)/\mathbb{Z}_N) = \mathbb{Z}_N$ discussed e.g. in 
\cite{'tHooft:1977hy,'tHooft:1979uj,deForcrand:2002vs}.

Since the lattice link variables gauge transform 
at different points $U_\mu(x) \to g^{\dagger}(x) U_\mu(x) g(x+\hat{\mu})$, 
$SU(N)/\mathbb{Z}_N$ invariance cannot be recovered from the local 
cancellation of the $\mathbb{Z}_N$ dependence in $g(x)$ as in the continuum 
and must be imposed directly on $U_\mu(x)$. As a consequence in adjoint 
theories regularized on the lattice it is by construction impossible to 
define observables transforming under the fundamental representation, i.e. 
sensitive to the $\mathbb{Z}_N$ center of the gauge group: their expectation 
value will vanish identically irrespective of the dynamics of the theory.
Therefore the symmetry breaking arguments for the deconfinement transition 
mentioned above cannot apply. It remains an open question whether a 
non-perturbative regularization of Yang-Mills theories allowing both 
$SU(N)/\mathbb{Z}_N$ invariance {\it and} non-vanishing fundamental 
observables can be defined.

In spite of all these interesting problems adjoint actions have 
not been intensively studied in the literature. For $N=2$ difficulties in 
their analysis have been well known for a long time 
\cite{Bhanot:1981eb,Greensite:1981hw,Halliday:1981te,Halliday:1981tm}:
the theory exhibits a bulk transition related to the
condensation of $\mathbb{Z}_2$ monopole charges $\sigma_c = -1$ which 
hinders the study
of its finite temperature properties. First concrete efforts to study the 
theory at finite temperature by implementing a
$\mathbb{Z}_2$ monopole suppressing chemical potential, as suggested in 
\cite{Halliday:1981te,Halliday:1981tm}, were made 
ten years ago
\cite{Cheluvaraja:1996zn,Datta:1997nv} reviving the interest in the subject. 
However, given the absence of a natural order parameter, attempts to locate a 
transition within phase II (the phase characterized by strong $\mathbb{Z}_2$ 
monopole suppression, see Fig.~\ref{paths}) through thermodynamic
observables were only conclusive in the strong coupling region
\cite{Datta:1999di,Datta:1999np}. In these works it was also first observed
how in some regions of phase II close to the bulk transition the theory 
possesses new states where the adjoint Polyakov loop $L_A \to -1/3$, 
additionally to the expected states where $L_A \to 1$. 
In Ref.~\cite{deForcrand:2002vs} a dynamical observable measuring the 
twist expectation value $z$, i.e. the topological index linked to 
$\pi_1(SU(N)/\mathbb{Z}_N)$, was introduced noting 
that the $\delta(\sigma_c-1)$ constraint effectively implemented by a
$\mathbb{Z}_2$ monopole suppression should
allow the $SO(3)$ partition function to be rewritten
as the sum of $SU(2)$
partition functions with all possible twisted boundary conditions 
$Z|_{z={\rm i}}$, ($i=0, \ldots, 3$ for $SU(2)$ on a 3+1 dimensional torus) 
\cite{Mack:1979gb,Tomboulis:1980vt,Kovacs:1998xm,Alexandru:2000wp}.
The $L_A \to -1/3$ state was thus linked directly to a 
non-trivial twist expectation value, equivalent to the creation of a vortex. 
Creating such 't~Hooft loop amounts to changing the
signs of some fundamental plaquettes, which however leaves
the adjoint action unmodified. This implies that $\Delta U=0$ in the 
free energy change $\Delta F=\Delta U-T \Delta S$, which 
will then only receive an entropy contribution. 
Defining thus the 't~Hooft vortex 
free energy $F/T = -\log{({Z|_{z=1}}/{Z|_{z=0}})}$ simply by the 
ratio of the partition function in the non-trivial twist sector to that in the 
trivial one, their relative weight being measured through an ergodic 
simulation, the $SO(3)$ theory was proposed as the ideal test case
to check whether the 't~Hooft vortex confinement criterion 
\cite{'tHooft:1977hy,'tHooft:1979uj} could compensate for the absence of an 
explicit order parameter linked to center symmetry breaking: in the 
thermodynamic limit ($V=N_s^3\to\infty$)
$F$ should vanish in the confined phase while diverging with an area law 
$F\sim \tilde{\sigma}N_s^{2}$ above the deconfinement transition. 
Working without the monopole suppression term proved however
to be a hurdle, since the ``freezing'' of twist sectors above the bulk 
transition creates high potential barriers hard to overcome even with a 
multi-canonical algorithm \cite{deForcrand:2002vs}, making ergodic 
simulations on top of the bulk transition unviable already for volumes 
larger than $8^3 \times 4$. 
Furthermore, since one would eventually need to go well beyond 
the bulk transition deeply into phase II with the simulations, 
the suitability of multihistogram \cite{Swendsen:1987ce}
or multicanonical methods \cite{Berg:1991cf} remains dubious. Ergodicity
problems and non-trivial twist sectors were not considered in 
Ref.~\cite{Datta:1999di,Datta:1999np}.

A particular observation has proven crucial in our taming of the
tunnelling problem: the bulk transition weakens with increasing 
$\mathbb{Z}_2$ monopole suppression, eventually becoming 2$^{\rm nd}$
order at some intermediate point \cite{Datta:1999di}.
Through the twist susceptibility the 2$^{\rm nd}$ order branch of
the bulk transition was shown to be consistent with the
4-d Ising model universality class
\cite{Datta:1999di,Barresi:2001dt,Barresi:2002un,Barresi:2006gq},
as expected by theoretical arguments
\cite{Halliday:1981te,Halliday:1981tm}.
To actually pin down the point where the transition changes from
weak first to second order is a difficult numerical task. This however 
has no practical consequences, since for the following it is
immaterial whether one deals with a second or a very weak first order bulk.

Although tunnelling among topological sectors is still suppressed with a 
local update algorithm, twists were shown to be well defined throughout 
phase II. $L_A$ on the other hand approximately satisfies 
\cite{Smilga:1993vb,Michael:1985ne} 
a Haar-measure distribution for low $\beta_A$, departing from it above 
some $\beta_A^c$. The critical value $\beta_A^c$ was seen to 
scale properly with the lattice extent in the Euclidean time direction
$N_{\tau}$ \cite{Barresi:2003jq,Barresi:2006gq}. This hints at a transition 
line (the dashed horizontal line in Fig.~\ref{paths})
separating a confining from a deconfining phase 
in each {\it fixed} twist sector 
\cite{Barresi:2003jq,Barresi:2004gk,Barresi:2003yb,Barresi:2006gq} 
collapsing on the bulk transition for the
$N_{\tau}$ commonly used in simulations.
It is therefore sound to conjecture that the whole physically relevant
$SO(3)$ dynamics lies in phase II, the finite temperature transition 
eventually decoupling from the bulk transition for high enough $N_\tau$ 
even without monopole suppression term. Unfortunately, according to
estimates in Ref.~\cite{deForcrand:2002vs} 
this should not happen for lattice sizes 
smaller than $\sim 800^3 \times 400$. A non-vanishing
$\mathbb{Z}_2$ monopole chemical potential together with an ergodic 
algorithm suitable for simulations throughout phase II seems therefore the 
only feasible way to gain access to the properties of the continuum limit of 
$SO(3)$.

Given the failure of center symmetry breaking criteria to identify the 
deconfinement transition in the adjoint theory, in 
\cite{Barresi:2004qa,Barresi:2006gq} 
the use of the Pisa disorder parameter for monopole condensation was proposed. 
Lines of second order transition properly scaling with $N_\tau$ and ending on 
the bulk transition line where actually found at {\em fixed} twist, with 
critical exponents consistent with the 3-d Ising model. Whether this 
is the case also for the ergodic theory, i.e. summed over all 
twist sectors, is the subject of the present paper. We will employ parallel 
tempering and utilize the mentioned weakening of the bulk transition
to overcome the high potential barriers that prevent tunnelling 
with local update algorithms. Moreover, ergodicity being 
an essential prerequisite for an unbiased measure of the vortex
free energy, it is an interesting question in its own right
whether such observable could indeed also play the r\^ole of an
order parameter for the deconfinement transition in $SO(3)$ 
\cite{deForcrand:2002vs,deForcrand:2001nd}. We will extend here the discussion
of the vortex free energy we have recently published in 
Ref.~\cite{Burgio:2006dc}.
Some preliminary results of the present project were also presented in 
\cite{Burgio:2005xe}. 

\section{The setup and the observables}
\label{sec2}
%--------------------------------------
As anticipated, we will concentrate on the adjoint $SU(2)$ Wilson action 
modified by a $~\mathbb{Z}_2$ monopole suppression term
\begin{eqnarray} 
S=\beta_{A} \sum_{P} 
  \left(1-\frac{1}{3}\mathrm{Tr}_A U_{P}\right)
  +\lambda \sum_{c}(1-\sigma_{c})\,,
\label{ouraction}
\end{eqnarray}
where $U_P$ denotes the standard plaquette variable and 
$\mathrm{Tr}_A O= (\mathrm{Tr}_F O)^2-1 = \mathrm{Tr}_F (O^2)+1$ the adjoint 
trace. The center blind product
$\sigma_{c}=\prod_{P\in\partial c}\mathrm{sign}(\mathrm{Tr}_{F}U_{P})$
taken around elementary 3-cubes $c$ defines the 
$\mathbb{Z}_2$ magnetic charge. Its density 
$M = 1-\langle\frac{1}{N_c}\sum_{c}\sigma_{c}\rangle$
tends to unity in the strong coupling region (phase I)
and to zero in the weak coupling limit (phase II), $N_c$ denoting the total 
number of elementary 3-cubes.  
The corresponding path-integral quantized lattice theory with the 
action~(\ref{ouraction}) is center-blind
in the entire $\beta_A-\lambda$ plane \cite{Barresi:2003jq}. 

The Pisa disorder parameter $\mu$
\cite{DiGiacomo:1997sm,DiGiacomo:1999fa,DiGiacomo:1999fb,Carmona:2001ja}
has been introduced for action~(\ref{ouraction}) in 
Ref.~\cite{Barresi:2004qa}. 
Its expectation value is given by 
$\langle\mu\rangle=\langle e^{- \Delta S}\rangle$, 
where 
$\Delta S=S^{M}-S$ is the difference of the standard plaquette action 
$S$ and an action $S^{M}$ modified by the introduction of an adjoint
bosonic field transforming at the space boundary under $G\sim SU(2)/U(1)$ 
\cite{DiGiacomo:1997sm,Frohlich:2000zp}.
Its evaluation does not require any gauge fixing, a point of view we adopt
in what follows \cite{Frohlich:2000zp}. 
We want to stress here that the introduction of $C^{*}$ boundary
conditions in the temporal direction, necessary to conserve magnetic
charge when defining $S^{M}$ at finite temperature, poses no conceptual 
problem in the adjoint theory, being equivalent up to a gauge rotation 
to a partial twist, i.e. only in the time direction \cite{Barresi:2004qa}. 
Since our adjoint action with periodic boundary conditions allows all twist 
matrices to be generated dynamically in any direction \cite{deForcrand:2002vs},
$C^{*}$ boundary conditions will just amount to a relabelling of the twist 
sectors. We will come back to this point later on.

Appropriate twist variables are introduced by \cite{deForcrand:2002vs}
\begin{eqnarray}                                          
z_{\mu\nu}\equiv\frac{1}{N_\rho N_\sigma}\sum_{\rho\sigma}
                \prod_{P\; \in\; \mathrm{plane} \; \mu\nu} \!\!
\mathrm{sign Tr}_F U_P\,,                                 
\hspace{.1cm}(\epsilon_{\rho\sigma\mu\nu}=1) .            
\label{deftwi}                                            
\end{eqnarray}                                            
Since the temporal twists in the various spatial directions 
$z_{i,4}\,,i=1,2,3$ are well identified (either +1 or -1) for
each configuration in phase II,
the partition functions restricted to a fixed
twist sector are easy to define as expectation values of suitable projectors 
\cite{'tHooft:1979uj}. Explicitly we have
\begin{eqnarray}
{\frac{Z|_{z=0}}{Z}}=\langle \nu_0 \rangle,\;\;\;&\nu_0& ={\frac{1}{8}}
\prod_{i=1}^3\lbrack1+\mathrm{sign}(z_{i,4})\rbrack\nonumber\\
{\frac{Z|_{z=1}}{Z}}=\langle \nu_1 \rangle,\;\;\;&\nu_1& ={\frac{1}{8}}
\sum_{j=1}^3\;\prod_{i=1}^3 \lbrack1+(1-2\delta_{i,j})
\mathrm{sign}(z_{i,4})\rbrack\nonumber\\              
{\frac{Z|_{z=2}}{Z}}=\langle \nu_2 \rangle,\;\;\;&\nu_2& ={\frac{1}{8}}
\sum_{j=1}^3\;\prod_{i=1}^3 \lbrack1-(1-2\delta_{i,j})
\mathrm{sign}(z_{i,4})\rbrack\nonumber\\              
{\frac{Z|_{z=3}}{Z}}=\langle \nu_3 \rangle,\;\;\;&\nu_3& ={\frac{1}{8}}
\prod_{i=1}^3\lbrack1-\mathrm{sign}(z_{i,4})\rbrack   
\label{fractions}
\end{eqnarray}
$\nu_k$ being equal to unity if the configuration belongs to the 
$k^{\rm th}$ sector and vanishing otherwise.

From Eq.~(\ref{fractions}) it follows that  
\begin{equation}
F= -T \log{ {\frac{Z_1}{3 Z_0}}}=
-{\frac{1}{a N_\tau}} \log{ {\frac{\langle\nu_1\rangle}
{3 \langle\nu_0\rangle}}}\,.
\label{free_en}
\end{equation}
The factor in the 
denominator is due to the three equivalent ways to measure $z_{i,4}=- 1$
on $\mathbb{T}^3\times\mathbb{S}$, rather than one as on 
$\mathbb{S}^3\times\mathbb{S}$; in this way $F$ will be normalized to zero 
if 0- and 1-twists are 
equally probable. This occurs on top of the bulk transition and in some sense 
everywhere in phase I, where twist sectors are however badly defined, because 
of the twist variables (\ref{deftwi}) 
fluctuating around zero.

\begin{figure}[thb]
\begin{center}
\includegraphics[angle=0,width=0.5\textwidth]{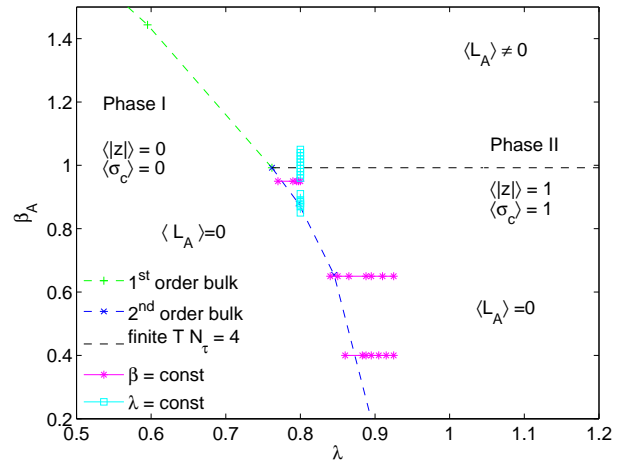}
\end{center}
\caption{Paths chosen for main simulations in the $\lambda-\beta_A$ plane.}
\label{paths}
\end{figure}

We will employ parallel tempering to achieve
ergodicity over different twist sectors when evaluating the expectation
values of physical observables, e.g. the Pisa disorder parameter and the
't~Hooft vortex free energy. Simulations have been 
carried out along the paths shown in Fig.~\ref{paths}.
The motivation for these choices will become clear in the following. 
Our spatial lattice sizes will vary between $N_s=12$ and $N_s=24$. 
The time-like extension will remain fixed ($N_{\tau}=4$).

\section{Parallel tempering}
%---------------------------
\subsection{General description}
%-------------------------------
In tempering methods some parameters of the action are made dynamical variables
in the simulations, updating the system in an enlarged configuration 
space. This allows a detour in parameter space  if a high tunnelling barrier 
is present at some parameter value, resulting in an improved algorithm. 

In the method of \emph{simulated} tempering first proposed in 
\cite{Marinari:1992qd} the inverse temperature is made a dynamical variable. 
With such algorithms considerable improvements have been obtained when 
rendering dynamical e.g. the number 
of degrees of freedom in the Potts-Model \cite{Kerler:1992ks}, the inverse 
temperature for spin glass \cite{Kerler:1994} and the monopole 
coupling in U(1) lattice theory \cite{Kerler:1994qc,Kerler:1995nj}. With 
dynamical mass of 
staggered fermions in full QCD \cite{Boyd:1996ww} a better sampling
of the configuration space has been reported. However, simulated tempering 
requires the determination of a weight function in the generalized action, 
and an efficient method of estimating it 
\cite{Kerler:1994,Kerler:1994qc,Kerler:1995nj} is crucial
for successfully accelerating the simulation.

A major progress was the proposal of the \emph{parallel} tempering method (PT)
\cite{Hukushima:1995,Marinari:1996dh}, in which no weight function needs to be 
determined. This method 
has allowed great improvements for spin glasses 
\cite{Hukushima:1995}. In QCD 
with dynamical quark mass better sampling has been reported for staggered 
fermions \cite{Boyd:1997rb}. In simulations of QCD with O($a$)-improved Wilson 
fermions \cite{Joo:1998ib} no computational advantage has been found when
making only 
two (relatively small) hopping parameter values dynamical. In subsequent 
works \cite{Ilgenfritz:1999hy,Ilgenfritz:2000nj} with more ensembles and 
standard Wilson fermions 
a considerable increase of the transitions between topological sectors has
been observed. In Ref.~\cite{Ilgenfritz:2001jp} these investigations 
have been extended to a detailed comparison 
with conventional simulations. Unfortunately no gain could be confirmed in
that case due to the region of parameter space 
used in which the mechanism of an easier detour was not available.

In the present application the fact that above the bulk phase transition
the barriers between the twist sectors cannot be overcome at all by 
conventional 
algorithm makes PT in any case superior. With a chain of parameter
points crossing the transition line along the softer branch of the
bulk transition the idea of an easier detour by tempering
is ideally realized. This is also reflected by the remarkably good efficiency
of PT observed.

\subsection{Parallel tempering algorithm}
%----------------------------------------

In standard Monte Carlo simulations one deals with one parameter set $p$ 
and generates a sequence of field configurations ${\cal_{F}}(s)$, where $s$ 
denotes the Monte Carlo time. In our case $p$ will include the coupling 
$\beta_A$ and the chemical potential $\lambda$.
In parallel tempering (PT) \cite{Hukushima:1995,Marinari:1996dh} 
one updates $K$ field 
configurations ${\cal_{F}}_{n}$ with $n = 1, \dots, K$ in the same run. The 
characteristic feature is that the assignment of the parameter sets 
$p_j$ with $j = 1, \dots, K$ to the field configurations ${\cal_{F}}_{n}$ 
changes in the course of a tempered simulation. The global configuration at 
time $s$ will be denoted by $B(s)$, ${\cal_{F}}_1(s)$, 
${\cal_{F}}_2(s)$,..., ${\cal_{F}}_K(s)$ where 
the permutation 
\begin{equation}
B(s)=\left(\begin{array} 
{cccccc}n_1(s)&n_2(s)&\ldots&n_j(s)&\ldots&n_K(s)\\
           1    &    2   &\ldots&    j   &\ldots&    K \\
\end{array}\right)
\label{eq:permutation}
\end{equation}
describes the assignment of the field configurations ${\cal_{F}}_{n_j(s)}(s)$
to the parameter sets $p_{j}$.  For short this approach is called PT 
with $K$ ensembles.

The update of the ${\cal_{F}}_{n}$ is implemented through a standard 
Metropolis procedure using the parameter sets $p_j$ as assigned at a 
given time. The update of $B$ is achieved by swapping pairs according to 
a further Metropolis acceptance condition with probability
\begin{equation}
P_{\rm swap}(i,j) = \min\left( 1, e^{-\Delta S} \right) \,,
\end{equation}
where the variation
\begin{eqnarray} \nonumber
\Delta S = 
&+& S(p_i, {\cal_{F}}_{n_i}) + S(p_j, {\cal_{F}}_{n_j}) \\
&-& S(p_i, {\cal_{F}}_{n_j}) - S(p_j, {\cal_{F}}_{n_i}) 
\label{eq:Pswap-1}
\end{eqnarray}
refers to the action $S$ for the parameter set $p_j$ and 
the field configurations ${\cal_{F}}_{n_j}$.
The total update of the Monte Carlo 
algorithm, after which its time $s$ increases by one, then consists of 
the updates of all ${\cal_{F}}_m$ followed by the full update of $B$ 
with a sequence of attempts to swap pairs.

Detailed balance for the swapping follows from Eq.~(\ref{eq:Pswap-1}). 
Ergodicity is obtained by updating all ${\cal_{F}}_{n}$ and by swapping 
pairs in 
such a way that all permutations of Eq.~(\ref{eq:permutation}) can be reached. 
There remains still the freedom of choosing the succession of the individual 
steps. Our choice is such that the updates of all ${\cal_{F}}_{n}$ and 
that of $B$ alternate. Our criterion for choosing the succession of swapping 
pairs in the update of $B$ has been to minimize the average time it takes 
for the assignment of a field configuration to the parameters to travel 
from the first to the last pair of parameter values. This
has led us to swap neighboring pairs and to
proceed with this along the respective path in Fig.~\ref{paths}. 

Observables of interest, associated to a specific set 
$p_j$, will be denoted as
\begin{equation}
\mathcal{O}_j(s) \equiv \mathcal{O}({\cal_{F}}_{n_j(s)}(s)),
\quad j=1,\ldots,K \,.
\label{eq:observables}
\end{equation}
As anticipated above, for the success of the method the softening of the 
bulk transition is crucial, 
since we need to ``transport'' the tunnelling that occurs in phase I
and on top of the bulk transition 
into phase II, where twist sectors are well defined but frozen. 
To work at low $\lambda$, i.e. on top of a strong 1$^{\rm st}$ 
order bulk, would select too high barriers and kill any 
hope of ergodicity at large
volume, as experienced in Ref.~\cite{deForcrand:2002vs,deForcrand:2001nd}. 
Moreover, in that parameter range for lattice sizes reasonably available to
the simulations (i.e. $\ll 800^3 \times 400$!) finite volume effects would 
still cover the physical transition \cite{deForcrand:2002vs}.

Some care is of course necessary also with our method. 
In particular to maintain a sufficient swapping acceptance rate $\omega$, 
i.e. to avoid the freezing of twist sectors, the 
distance between neighboring couplings must diminish with the volume. 
On the other hand to keep cross-correlations 
under control one does not wish the acceptance rate to be too high. We
have chosen to tune the parameters for each path and volume at hand
so to keep the acceptance rate roughly fixed at around $\omega = 12 \%$, 
a value for which we empirically find a good balance between auto- and
cross-correlations. We also find that performing some standard Metropolis
overupdate hits on the ${\cal_{F}}_n$ before the actual PT update 
(\ref{eq:Pswap-1}) is proposed helps in diminishing correlations.

The relatively low value of $\omega$ has an intuitive explanation:
for each parameter set $p_j$ one wishes to ``explore'' the various
twist sectors for a sufficiently long MC time before tunnelling.
It causes however also a technical problem: if the starting configurations are
all in the same twist sector the ensemble needs a very long time before the
``disorder'' below the bulk spreads to the configurations further above it.
An efficient way out is to randomly choose the twist sectors of the elements
in the start ensemble.

%-------------------
\begin{table}[thb]
\begin{center}
\begin{tabular}{||c|c||c|c||c|c|c|c||}
 \hline\hline
 \multicolumn{2}{||c||}{$N_s = 12$} & \multicolumn{2}{c||}{$N_s = 16$} & 
 \multicolumn{2}{c||}{$N_s = 20$(a)}& \multicolumn{2}{c||}{$N_s = 20$(b)}\\ 
 \hline
 \multicolumn{2}{||c||}{$2\times40000$} & 
 \multicolumn{2}{c||}{$2\times40000$} & 
 \multicolumn{2}{c||}{$2\times40000$}& 
 \multicolumn{2}{c||}{$2\times40000$}\\ 
 \hline\hline
 $\lambda$& $\beta_A$& $\lambda$& $\beta_A$& $\lambda$& $\beta_A$& 
 $\lambda$& $\beta_A$\\
        \hline
        0.78 & 0.960 & 0.78 & 0.960 & 0.77  & 0.960  & 0.77  & 0.960  \\
        0.79 & 0.960 & 0.79 & 0.960 & 0.78  & 0.960  & 0.78  & 0.960  \\
        0.80 & 0.960 & 0.80 & 0.960 & 0.79  & 0.960  & 0.79  & 0.960  \\
        0.80 & 0.975 & 0.80 & 0.970 & 0.80  & 0.960  & 0.80  & 0.960  \\
        0.80 & 0.990 & 0.80 & 0.981 & 0.80  & 0.970  & 0.80  & 0.970  \\
        0.80 & 1.005 & 0.80 & 0.993 & 0.80  & 0.980  & 0.80  & 0.980  \\
        0.80 & 1.020 & 0.80 & 1.006 & 0.80  & 0.990  & 0.80  & 0.990  \\
        0.80 & 1.035 & 0.80 & 1.019 & 0.80  & 1.000  & 0.80  & 1.005  \\
        0.80 & 1.050 & 0.80 & 1.032 & 0.80  & 1.010  & 0.80  & 1.015  \\
        0.80 & 1.065 & 0.80 & 1.045 & 0.80  & 1.025  & 0.80  & 1.035  \\
        0.80 & 1.080 & 0.80 & 1.058 & 0.80  & 1.040  & 0.80  & 1.045  \\
        0.80 & 1.090 & 0.80 & 1.070 & 0.80  & 1.050  & 0.80  & 1.055  \\
        \hline\hline
\end{tabular}
\end{center}
\caption{Lattice sizes, statistics and couplings for PT
         simulations in Fig.~\ref{allw} and Fig.~\ref{pivstw}, right branch.}
\label{tab1}
\end{table}
 
\begin{table}[thb]
\begin{center}
\begin{tabular}{||c|c||c|c||c|c||}
 \hline\hline
 \multicolumn{2}{||c||}{$N_s = 12$} & \multicolumn{2}{c||}{$N_s = 16$} &
 \multicolumn{2}{c||}{$N_s = 20$}\\ 
  \hline
 \multicolumn{2}{||c||}{$2\times40000$} & 
 \multicolumn{2}{c||}{$2\times40000$} &  
 \multicolumn{2}{c||}{$2\times40000$} \\ 
 \hline\hline
 $\lambda$& $\beta_A$& $\lambda$& $\beta_A$& $\lambda$& $\beta_A$\\
        \hline
        0.80 & 0.860 & 0.80 & 0.860 & 0.80  & 0.860  \\
        0.80 & 0.870 & 0.80 & 0.870 & 0.80  & 0.870  \\
        0.80 & 0.875 & 0.80 & 0.880 & 0.80  & 0.875  \\
        0.80 & 0.880 & 0.80 & 0.890 & 0.80  & 0.880  \\
        0.80 & 0.885 & 0.80 & 0.895 & 0.80  & 0.885  \\
        0.80 & 0.890 & 0.80 & 0.900 & 0.80  & 0.890  \\
        0.80 & 0.900 & 0.80 & 0.908 & 0.80  & 0.895  \\
        0.80 & 0.905 & 0.80 & 0.920 & 0.80  & 0.908  \\
        0.80 & 0.908 & 0.80 & 0.925 & 0.80  & 0.925  \\
        0.80 & 0.910 &      &       &       &        \\
        0.80 & 0.920 &      &       &       &        \\
        0.80 & 0.925 &      &       &       &        \\
        \hline\hline
\end{tabular}
\end{center}
\caption{Parameter sets for PT runs in Fig.~\ref{pivstw}, 
         left branch.}
\label{tab2}
\end{table}

\begin{table}[thb]
\begin{center}
\begin{tabular}{||c|c||c|c||c|c||}
  \hline\hline
  \multicolumn{2}{||c||}{$N_s = 12$} & \multicolumn{2}{c||}{$N_s = 16$} & 
  \multicolumn{2}{c||}{$N_s = 20$}\\ 
 \hline
 \multicolumn{2}{||c||}{$30000$} & 
 \multicolumn{2}{c||}{$30000$} &  
 \multicolumn{2}{c||}{$30000$} \\ 
 \hline\hline
 $\lambda$& $\beta_A$& $\lambda$& $\beta_A$& $\lambda$& $\beta_A$\\
 \hline
        0.80 & 0.865 & 0.80 & 0.865 & 0.80  & 0.865  \\
        0.80 & 0.870 & 0.80 & 0.870 & 0.80  & 0.870  \\
        0.80 & 0.875 & 0.80 & 0.875 & 0.80  & 0.875  \\
        0.80 & 0.880 & 0.80 & 0.880 & 0.80  & 0.880  \\
        0.80 & 0.890 & 0.80 & 0.890 & 0.80  & 0.890  \\
        0.80 & 0.900 & 0.80 & 0.900 & 0.80  & 0.900  \\
        0.80 & 0.910 & 0.80 & 0.910 & 0.80  & 0.910  \\
        0.80 & 0.920 & 0.80 & 0.920 & 0.80  & 0.920  \\
        0.80 & 0.930 & 0.80 & 0.930 & 0.80  & 0.930  \\
        0.80 & 0.935 & 0.80 & 0.935 & 0.80  & 0.935  \\
 \hline\hline
\end{tabular}
\end{center}
\caption{Parameters sets for PT runs
             in Fig.~\ref{free1}, left branch.}
\label{tab3}
\end{table}

\begin{table}[thb]
\begin{center}
\begin{tabular}{||c|c||c|c||c|c||c|c||}
 \hline\hline
 \multicolumn{2}{||c||}{$N_s = 12$} & \multicolumn{2}{c||}{$N_s = 16$} & 
 \multicolumn{2}{c|}{$N_s = 20$}& \multicolumn{2}{c||}{$N_s = 24$} \\ 
 \hline
 \multicolumn{2}{||c||}{$100000$} & 
 \multicolumn{2}{c||}{$100000$} & 
 \multicolumn{2}{c||}{$100000$}& 
 \multicolumn{2}{c||}{$100000$}\\ 
 \hline\hline
 $\lambda$& $\beta_A$& $\lambda$& $\beta_A$& $\lambda$& 
 $\beta_A$& $\lambda$& $\beta_A$\\
        \hline
        0.78 & 0.95 & 0.78 & 0.95 & 0.785 & 0.95  & 0.785 & 0.95  \\
        0.79 & 0.95 & 0.79 & 0.95 & 0.795 & 0.95  & 0.795 & 0.95  \\
        0.795& 0.95 & 0.795& 0.95 & 0.7975& 0.95  & 0.7975& 0.95  \\
        0.80 & 0.95 & 0.80 & 0.95 & 0.80  & 0.95  & 0.80  & 0.955 \\
        0.80 & 0.96 & 0.80 & 0.96 & 0.80  & 0.96  & 0.80  & 0.960 \\
        0.80 & 0.97 & 0.80 & 0.97 & 0.80  & 0.967 & 0.80  & 0.966 \\
        0.80 & 0.98 & 0.80 & 0.98 & 0.80  & 0.974 & 0.80  & 0.972 \\
        0.80 & 0.99 & 0.80 & 0.99 & 0.80  & 0.981 & 0.80  & 0.978 \\
        0.80 & 1.00 & 0.80 & 1.00 & 0.80  & 0.988 & 0.80  & 0.984 \\
        0.80 & 1.01 & 0.80 & 1.01 & 0.80  & 0.995 & 0.80  & 0.991 \\
        0.80 & 1.02 & 0.80 & 1.02 & 0.80  & 1.002 &       &       \\
        0.80 & 1.03 & 0.80 & 1.03 & 0.80  & 1.009 &       &       \\
        0.80 & 1.04 & 0.80 & 1.04 & 0.80  & 1.016 &       &       \\
        0.80 & 1.05 & 0.80 & 1.05 & 0.80  & 1.023 &       &       \\
 \hline\hline
\end{tabular}
\end{center}
\caption{Parameters sets for PT runs
         in Fig.~\ref{free1}, right branch.}
\label{tab4}
\end{table}

\begin{table}[thb]
\begin{center}
\begin{tabular}{||c|c||c|c||c|c||}
 \hline\hline
 \multicolumn{2}{||c||}{$N_s = 12$} & \multicolumn{2}{c||}{$N_s = 16$} & 
 \multicolumn{2}{c||}{$N_s = 20$}\\ 
 \hline
 \multicolumn{2}{||c||}{$100000$} & 
 \multicolumn{2}{c||}{$100000$} & 
 \multicolumn{2}{c||}{$100000$}\\ 
 \hline\hline
 $\lambda$& $\beta_A$& $\lambda$& $\beta_A$& $\lambda$& $\beta_A$\\
 \hline
        0.80 & 0.92 & 0.80 & 0.92 & 0.80  & 0.932  \\
        0.80 & 0.93 & 0.80 & 0.93 & 0.80  & 0.939  \\
        0.80 & 0.94 & 0.80 & 0.94 & 0.80  & 0.946  \\
        0.80 & 0.95 & 0.80 & 0.95 & 0.80  & 0.953  \\
        0.80 & 0.96 & 0.80 & 0.96 & 0.80  & 0.960  \\
        0.80 & 0.97 & 0.80 & 0.97 & 0.80  & 0.967  \\
        0.80 & 0.98 & 0.80 & 0.98 & 0.80  & 0.974  \\
        0.80 & 0.99 & 0.80 & 0.99 & 0.80  & 0.981  \\
        0.80 & 1.00 & 0.80 & 1.00 & 0.80  & 0.988  \\
        0.80 & 1.01 & 0.80 & 1.01 & 0.80  & 0.995  \\
        0.80 & 1.02 & 0.80 & 1.02 & 0.80  & 1.002  \\
        0.80 & 1.03 & 0.80 & 1.03 & 0.80  & 1.009  \\
        0.80 & 1.04 & 0.80 & 1.04 & 0.80  & 1.016  \\
        0.80 & 1.05 & 0.80 & 1.05 & 0.80  & 1.023  \\
 \hline\hline
\end{tabular}
\end{center}
\caption{Parameter sets for PT runs
         in Figs.~\ref{times} and \ref{back}.}
\label{tab5}
\end{table}

\begin{table}[thb]
\begin{center}
\begin{tabular}{||c|c||c|c||}
 \hline\hline
 \multicolumn{2}{||c||}{$N_s = 16$} & \multicolumn{2}{c||}{$N_s = 16$}\\ 
  \hline
 \multicolumn{2}{||c||}{$30000$} & \multicolumn{2}{c||}{$30000$} \\ 
 \hline\hline
 $\lambda$& $\beta_A$& $\lambda$& $\beta_A$\\
        \hline
        0.870 & 0.40 & 0.830 & 0.65  \\
        0.883 & 0.40 & 0.850 & 0.65  \\
        0.888 & 0.40 & 0.865 & 0.65  \\
        0.895 & 0.40 & 0.888 & 0.65  \\
        0.905 & 0.40 & 0.895 & 0.65  \\
        0.915 & 0.40 & 0.910 & 0.65  \\
        0.925 & 0.40 & 0.925 & 0.65  \\
        \hline\hline
\end{tabular}
\end{center}
\caption{Parameter sets for two of the parallel tempering runs in 
Fig.~\ref{low}.}
\label{tab6}
\end{table}
%-------------------
Details for the investigated lattice sizes, the chosen parameter sets 
$~(\lambda, \beta_A)_j, ~j=1,2, \ldots, K~$ and the statistics for each 
ensemble for the paths drawn in Fig.~\ref{paths} 
are listed in the columns of Tables~\ref{tab1} to \ref{tab6}. 
Remember that the path along which we are passing through the 
finite temperature transition at fixed $~\lambda=0.8~$ starts 
with a horizontal piece at fixed $\beta_A=0.95$. 
The factor 2 for $N$ in Tables~\ref{tab1} and \ref{tab2} refers to the runs 
with and without modified action $S^M$, respectively. In order to
remain on the safe side in some results we have omitted the first and
last elements of the ensembles, since the latter have no further 
configuration to swap with. The respective errors of these points 
might not be of a comparable quality.
In the literature one can find that by adjusting the parameter spacing such 
that the endpoints get visited with the same probability as the neighboring
points the errors tend to become comparable. 

For the fixed $\lambda$ path of Fig.~\ref{paths}, along which the main
simulations have been performed, we can fit very well
the step $~\delta \beta_A~$ needed to keep $~\omega~$ fixed with a scaling 
law of the form
\begin{equation}
\delta \beta_{A}(\omega,N_s) \simeq \frac{\alpha(\omega)}{N_s^2}\,,
\label{scale1}
\end{equation}
where we find $\alpha(12\%)=2.15(3)$ in the $\beta_A = 0.95-1.09$ range 
considered, although we expect it to change with the width and location
of the the $(\lambda,\beta_A)$ window.
Being $\omega$ nothing but the tunneling probability among twist sectors, 
it should be proportional to the probability to create a vortex. Since
the cost to generate the latter should scale with an area law 
\cite{'tHooft:1979uj}, the $N_s^2$ dependance of the former is easily 
understood. Eq.~(\ref{scale1}) implies that to explore a fixed region 
$\Delta \beta_A$ of parameter space the number of ensembles will scale like
\begin{equation}
K \simeq \frac{\Delta \beta_A}{\alpha} N_s^2
\label{scale}
\end{equation}

\begin{figure}[thb]
\begin{center}
\subfigure{\includegraphics[angle=0,width=0.45\textwidth]{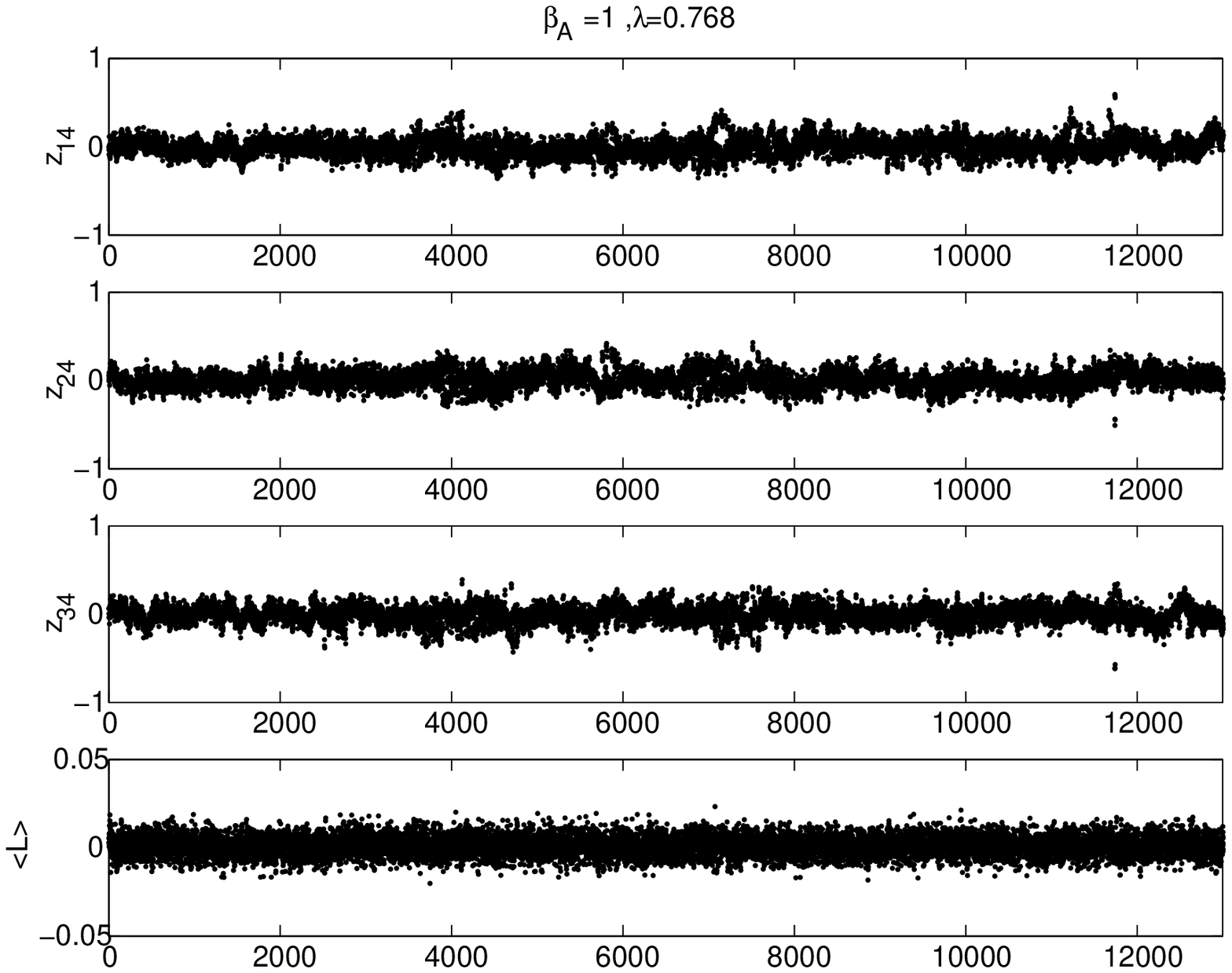}}
\subfigure{\includegraphics[angle=0,width=0.45\textwidth]{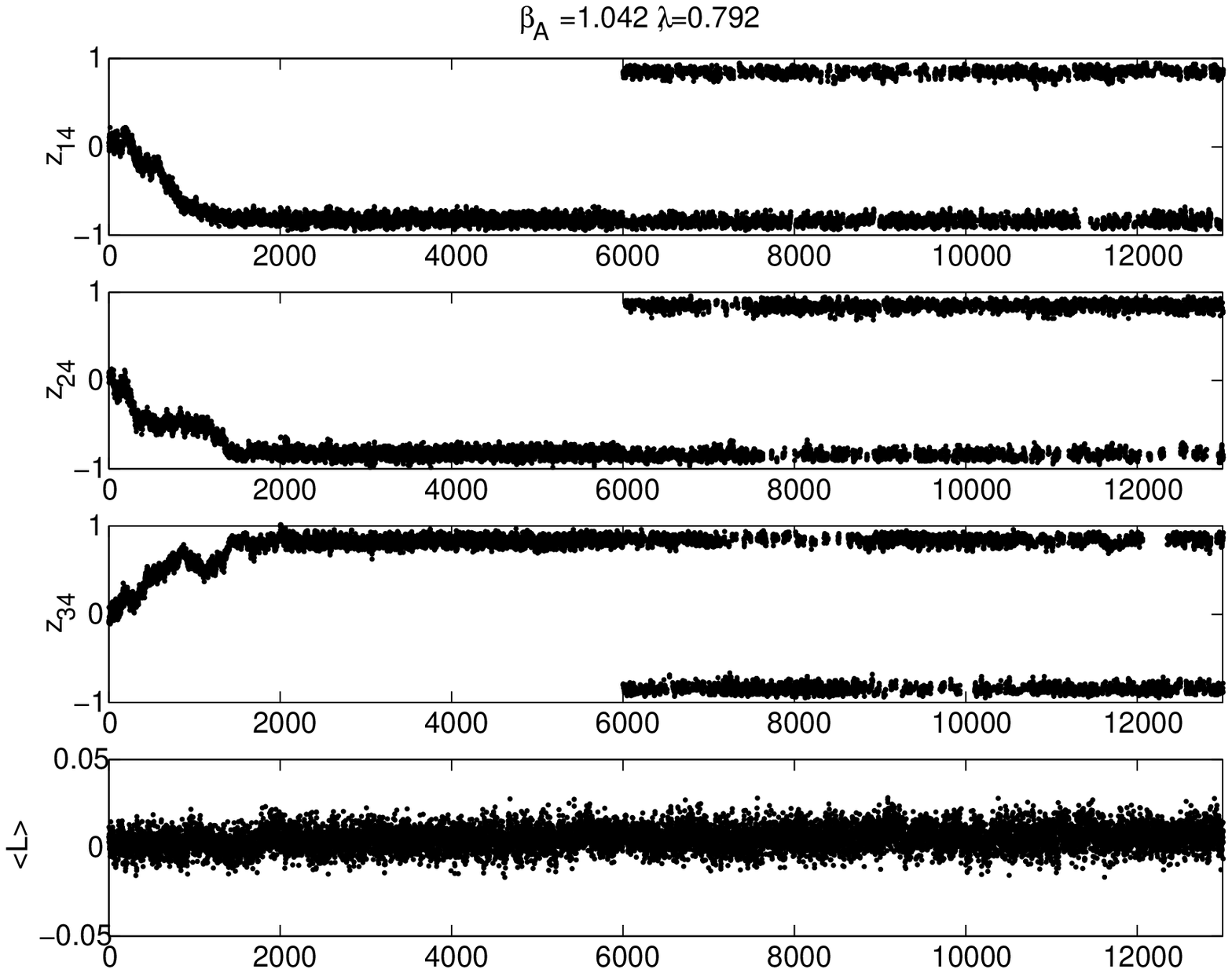}}
\end{center}
\caption{MC histories of twist variables $~z_{i,4},~i=1,2,3~$ and of the 
adjoint Polyakov loop $~L_A~$ for two PT ensembles. The lattice size is
$16^3 \times 4$.}
\label{history}
\end{figure}

As an illustration of the ergodicity of the algorithm we show in 
Fig.~\ref{history} the MC time histories of the twist observables 
$z_{i,4}$ and of the adjoint Polyakov loop for two ensembles 
belonging to the same test PT simulation with $K=10$ and $N_s =16$, 
one below (phase I) and one above the 
bulk transition (phase II). We have first let the single ensembles 
evolve separately with standard Metropolis updates, i.e. not
updating the permutation table $B$. The barriers among sectors are huge 
\cite{deForcrand:2002vs} and practically impossible to overcome without PT 
within phase II: the system is simply stuck in a fixed topological sector. 
Indeed, as Fig.~\ref{history} shows the twist variables remained stable 
over 6000 sweeps until we have turned on the full PT updates. The 
system started then frequently to tunnel among all sectors. Actually below 
the bulk transition there is no substantial difference between the two 
algorithms, since the disorder induced by the $\mathbb{Z}_2$ monopoles lets 
any algorithm be ergodic for the simple reason that topological sectors
are ill-defined, all twist values fluctuating around zero. The difference is 
however dramatic above the bulk transition in phase II, where tunnelling
among well-defined topological sectors is enabled by the PT 
algorithm. 

\subsection{Cross and autocorrelations}
%---------------------------------------
In PT the $K$ ensembles are generated in a correlated way. Therefore, the
full non-diagonal covariance matrix for the observables has to be taken into 
account. The latter
is obtained from the general correlation functions which, for an 
observable $\mathcal{O}$ and a number $N$ of updates, are defined as
\begin{eqnarray}
R_{jk}(t)&=&\frac{1}{N}\sum_{s=1}^N \mathcal{O}_j(s) \mathcal{O}_k(s+t)
\nonumber\\
&-&\Big(\frac{1}{N}\sum_{s'=1}^N \mathcal{O}_j(s')\Big)
\Big(\frac{1}{N}\sum_{s''=1}^N \mathcal{O}_k(s'')\Big) \,.
\label{eq:R-function-1}
\end{eqnarray}
For $j = k$ they are the usual autocorrelation functions, while for
$j \ne k$ they describe cross correlations between different ensembles.

The covariance matrix is obtained \cite{Ilgenfritz:2001jp} by using the 
general correlation function of Eq.~(\ref{eq:R-function-1}) and 
generalizing the derivation in Ref.~\cite{Priestley:1989} for the case 
$j=k$, which gives
\begin{eqnarray} \nonumber
C_{jk} &=& \frac{1}{N} R_{jk}(0) \\ &+& \frac{1}{N}\sum_{t=1}^{N-1}
\Big(1-\frac{t}{N}\Big)\Big(R_{jk}(t)+R_{kj}(t)\Big)\,.
\label{eq:C-function}
\end{eqnarray}
The diagonal elements of Eq.~(\ref{eq:C-function}) are the variances of 
$\mathcal{O}_j$ usually written as 
\begin{equation}
\mbox{var}(\mathcal{O}_j)=\frac{R_{jj}(0)}{N}\;2\tau_j \; ,
\label{eq:variance}
\end{equation}
with the integrated autocorrelation times $\tau_j$
\begin{equation}
\tau_j=\frac{1}{2}+\sum_{t=1}^{N-1}\rho_j(t),\qquad 
\rho_j(t)=R_{jj}(t)/R_{jj}(0). 
\label{eq:tau-int}
\end{equation}
When evaluating $\tau_j$ according to
Eq.~(\ref{eq:tau-int}) in practical simulations the summation up to $N-1$  
makes no sense since $\rho_j(t)$ is buried in the Monte Carlo 
noise already for relatively small $t$. Therefore, it has been proposed 
\cite{Priestley:1989,Madras:1988ei} to sum up only to some smaller  value 
$M$ of $t$. 
However, in practice such procedure is not stable against the choice of 
$M$ and neglecting the rest is a bad approximation. The proposal to 
estimate the remainder
by an extrapolation based on the $t$-values $M$ 
and $M-1$ \cite{Wolff:1989gz} is still inaccurate in general. A more
satisfying procedure is to describe the rest by a fit function based on 
the (reliable) terms of Eq.~(\ref{eq:C-function}) for $t \le M$ and on the 
general knowledge about the Markov spectrum. This procedure has led to 
very good results results in other applications \cite{Kerler:1993bn}.

In order to apply the latter strategy to determine the off-diagonal 
entries in Eq.~(\ref{eq:C-function}) one has to study how spectral properties 
enter the problem. This is possible introducing an appropriate Hilbert space 
\cite{Madras:1988ei,Kerler:1993bp}. Working this out, in 
Ref.~\cite{Ilgenfritz:2001jp} the general 
representation
\begin{equation}
R_{jk}(t)=\sum_{r>1}a_{jkr}\gamma_r^t \quad\mbox{with}\quad |\gamma_r|<1
\label{eq:R-function-3}
\end{equation}
has been obtained, where only the coefficients $a_{jkr}$ depend on the 
particular pair of observables while the eigenvalues $\gamma_r$ 
are universal and characteristic for the simulation algorithm.
To explain the behaviour of the off-diagonal elements the approximate 
functional form
\begin{equation} 
R_{jk}(t) \approx \left \{\begin{array} {cl}\sum_{r>1}\tilde{a}_{jkr}
\gamma_r^t
&\mbox{for}\quad|j-k|\le t\\0&\mbox{for}\quad0\le t<|j-k| \end{array}\right\}
\label{eq:Wtc}
\end{equation}
has been derived \cite{Ilgenfritz:2001jp} for $j \neq k$, which indicates a 
maximum at $t=|j-k|$.

For the numerical evaluation of Eq.~(\ref{eq:C-function}) the method mentioned
above is to be used, generalizing it to the off-diagonal elements.
The fits in the noisy region can exploit the universality of the Markov 
spectrum and the fact that after some time only the slowest mode survives.
Of course, such evaluation is limited by the available statistics.
To calculate errors one has to account for the cross correlations between 
the ensembles. To be able to do this one has to rely on fits to the data. 
The respective fit method is well know from the treatment of indirect 
measurements (see e.g.~Ref.~\cite{Brandt:1976zc}). For the application to PT 
the details have been worked out in Ref.~\cite{Ilgenfritz:2001jp}. 
To obtain errors for the covariances one can generalize the derivations 
given in Ref.~\cite{Priestley:1989} for the diagonal case to calculate 
covariances of covariances from the $R_{jk}(t)$ data only. However, in 
practice one can hardly get enough statistics for this.

\subsection{Correlation results}
\begin{figure}[thb]
\begin{center}
\subfigure{\includegraphics[angle=0,width=0.5\textwidth]{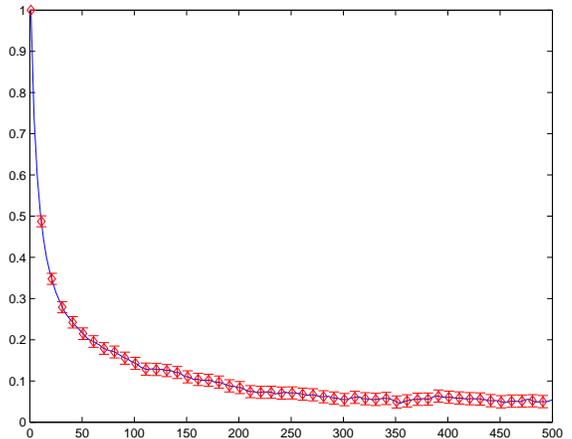}}
\end{center}
\caption{Example of a normalized diagonal correlation function for
$z_{1,4}$ 
at $\beta_A=0.98$ and $\lambda=0.8$ 
on a $16^3\times 4$ lattice.}
\label{acorr}
\end{figure}

\begin{figure}[thb]
\begin{center}
\subfigure{\includegraphics[angle=0,width=0.5\textwidth]{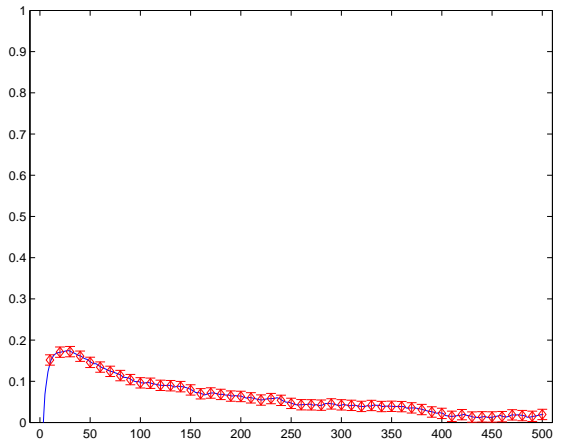}}
\end{center}
\caption{Example of a first off-diagonal normalized correlation function 
for $z_{1,4}$ 
between ensembles taken at $\beta_A=0.98$ and $0.99$ for $\lambda=0.8$
and lattice size $16^3\times 4$.}
\label{xcorr}
\end{figure}

\begin{figure}[thb]
\begin{center}
\subfigure{\includegraphics[angle=0,width=0.5\textwidth]{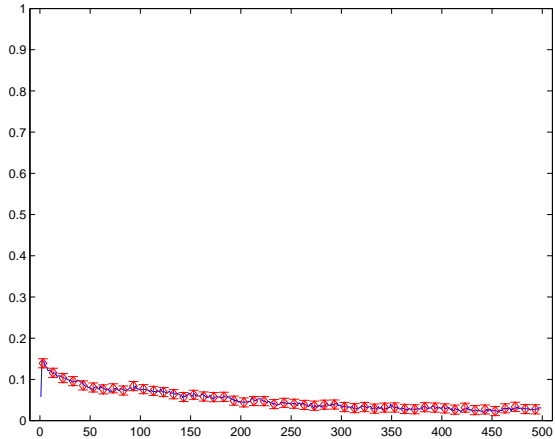}}
\end{center}
\caption{Example of a second off-diagonal normalized correlation function for 
$z_{1,4}$ 
between ensembles taken at $\beta_A=0.98$ and $1.00$ for $\lambda=0.8$
and lattice size on $16^3\times 4$.}
\label{xxcorr}
\end{figure}

%-------------------------------
Typical examples of correlation functions $R_{jk}(t)$ 
(normalized to $R_{jj}(0)$) are shown in 
Fig.~\ref{acorr},~\ref{xcorr} and \ref{xxcorr} for the twist variable 
$z_{1,4}$ as introduced in Section 2.
Along with all $z_{i,4}$ we have also used the autocorrelations for the
Polyakov loop as an additional source to determine the eigenvalues 
$\gamma_r$ in our analysis.
For the off-diagonal elements $R_{jk}(t)$ which show a clear
signal above the noise we generally observe a maximum at roughly $t=|j-k|$, 
thus seeing indeed the behaviour predicted by Eq.~(\ref{eq:Wtc}) 
(within errors) in our 
data. It is usually difficult to identify more than two or three 
off-diagonals above the noise. 
The correlations tend moreover to decrease with increasing volume. 
From our data we can clearly conclude that the off-diagonal 
elements of the general correlation functions are decreasing with the 
distance from the diagonal, their contributions being
reasonably smaller than the diagonal one, indicating that
cross correlations do not play an essential r\^ole. 

\begin{figure}[thb]
\begin{center}
\subfigure{\includegraphics[angle=0,width=0.5\textwidth]{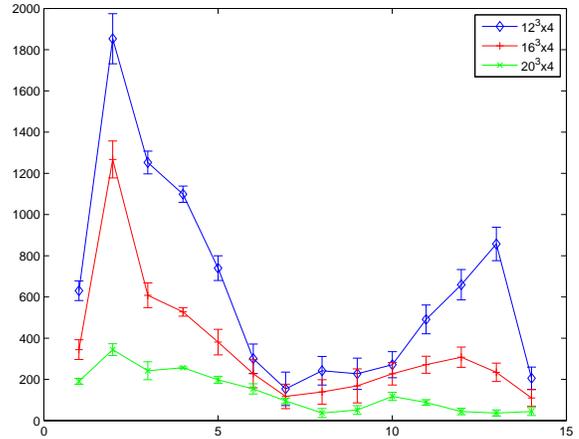}}
\end{center}
\caption{
Integrated autocorrelation times for the twist variable $z_{i4}$ 
at different volumes. The numbers at the horizontal axis enumerate the 
parameter points $(\lambda,\beta_A)$ for the PT ensembles corresponding 
to Tab. \ref{tab4}.}
\label{times}
\end{figure}

Fig.~\ref{times} shows the integrated autocorrelation times 
for all twist variables $~z_{i,4}$ obtained at different volumes
for each parallel configuration along the paths at $\lambda=0.8$ in
Fig.~\ref{paths}. 
Autocorrelations clearly decrease with the volume, as expected. As we shall 
see, the two peaks correspond to the bulk transition and the 
finite-temperature transition.

The observables that will suffer most from correlations in PT 
are obviously those whose expectation value depends significantly on
twist sectors. In our practical case only the twists themselves and 
the vortex free energy are sensitive to the ergodicity 
properties of the algorithm.
For such observables errors will be given by a combination of statistical
errors, estimated by bootstrapped sampling, and auto/cross-correlation
errors given by error propagation of the errors on $\nu_i$. 
Other observables, like the Polyakov loop and the Pisa disorder 
parameter, are roughly twist independent away from the deep deconfined 
phase, which we anyway do not reach in our simulations 
\cite{Barresi:2004qa,Barresi:2006gq}. For them only the statistical errors
will be relevant.

\section{Results}
%----------------

\subsection{Monopole condensation}
%---------------------------------
\label{mon_cond}

\begin{figure}[thb]
\begin{center}
\includegraphics[angle=0,width=0.5\textwidth]{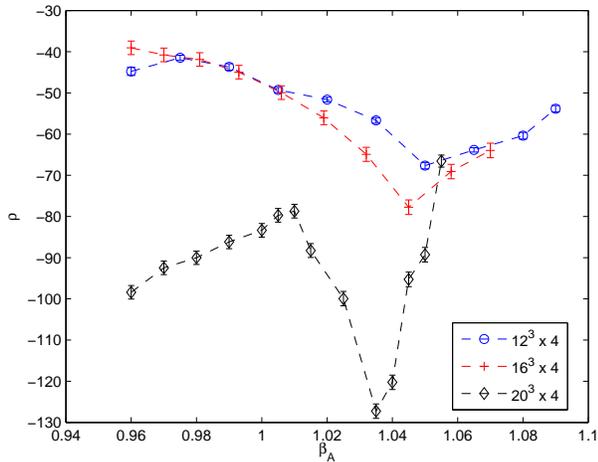}
\end{center}
\caption{$\rho$ vs. $\beta_A$ for $\lambda=0.8$ 
and various lattice sizes.}
\label{allw}
\end{figure}

The computation of the Pisa order parameter $\mu$ can be extended 
to the parallel tempering approach in a straightforward way.
Fig.~\ref{allw} shows $\rho=\frac{d}{d\beta_A}\log\langle\mu\rangle$ 
for fixed $\lambda=0.8$. As discussed in Ref.~\cite{Barresi:2004qa}, 
to prove confinement through monopole condensation $\rho$
should be small and bounded from below in the confined phase, 
display a dip 
at the deconfinement transition and reach a negative plateau whose value 
should scale like $- O(N_s \log{N_s})$ in the deep deconfined phase. 

\begin{figure}[thb]
\begin{center}
\includegraphics[angle=0,width=0.5\textwidth]{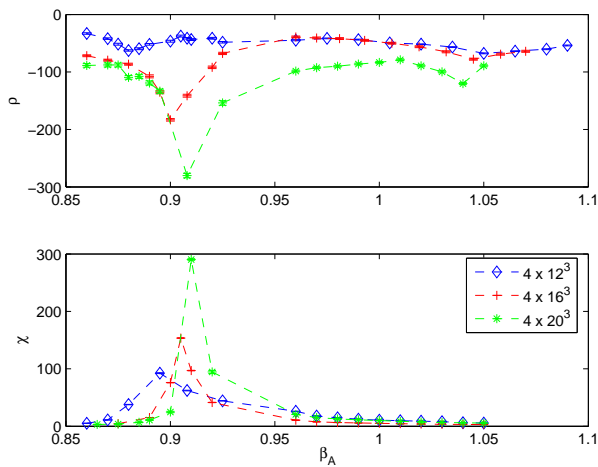}
\end{center}
\caption{Comparison between the Pisa order parameter $\rho$ (up) 
and the susceptibility of the average twist $\chi_z$ (below)
as a function of $\beta_A$ for $\lambda=0.8$.}
\label{pivstw}
\end{figure}

The dip in Fig.~\ref{allw} shows the position of the finite temperature 
transition. 
The region left to it, where $\rho$ should roughly vanish, 
is too close to the bulk transition to approach its $\beta_A \to 0$ value.
$\rho$ indeed has a dip at the bulk transition, as shown in 
Fig.~\ref{pivstw} \cite{Barresi:2004qa,Barresi:2006gq}, where
$\mathbb{Z}_2$ monopoles disappear.
Both phases left and right of the bulk 
transition are however still confining as long as $\rho$ 
remains bounded from below 
on both sides \cite{Barresi:2006gq}.
Fig.~\ref{pivstw} compares 
the occurrence of the second $\rho$-dip around $\beta_A \simeq 0.9$
with the existence of a corresponding peak in the susceptibility of the 
average twist as defined in \cite{Barresi:2003jq}. 
While the location of the latter peak is temperature-independent, its
height cannot be used to get the scaling with the 4-d volume since this
should be calculated at $T=0$. For the susceptibility the latter can be
done, obtaining critical exponents in accordance with Ising 4-d, as in 
\cite{Barresi:2003jq}. 
On the other hand, the Pisa disorder operator definition we use makes only 
sense at $T\neq 0$ (for a definition at $T=0$ see \cite{DiGiacomo:1997sm}).
Some caution is therefore necessary in interpreting the results of 
Fig.~\ref{allw}.
The region we investigate is, by the above exposed limitations 
of the algorithm, very close to both the bulk transition and the physical 
phase transition, so that two competing effects are superimposing. 
A thorough analysis of the whole phase space 
would be very expensive in terms of computer time and could
anyway hardly be extended to very high $\beta_A$. Nevertheless,
as argued in Ref.~\cite{Barresi:2006gq}, from
the fixed twist dynamics of our model \cite{Barresi:2004qa,Barresi:2006gq} one
can conclude that for the ergodic theory the Pisa disorder parameter
indicates condensation of monopoles
in the low $\beta_A$ region and deconfinement at high $\beta_A$, 
provided that a diverging dip at some $\beta_A^c$ exists, as
Fig.~\ref{allw} clearly shows. 

Indeed, given that the ergodic expectation value of $\mu$ can be written as
\begin{equation}
\langle\mu\rangle_{erg} = \frac{\sum_{\sf i} \mu|_{z={\sf i}} 
Z_{\sf SO(3)}|_{z={\sf i}}}{\sum_{\sf i} Z_{\sf SO(3)}|_{z={\sf i}}}\,,
\end{equation}
at large $\beta_A$, taking into account the observed 
$O(-N_s \log{N_s})$ plateaus of $\rho$ for trivial twist and its 
vanishing at non-trivial twist \cite{Barresi:2004qa,Barresi:2006gq} we have 
$\langle\mu\rangle_{erg} \simeq \langle\mu\rangle|_{z=0} (1-e^{-\frac{F}{T}})$.
The latter equation clearly implies an exponential vanishing of 
$\langle\mu\rangle_{erg}$ in the thermodynamic limit at high $\beta_A$.

At low $\beta_A$ one actually needs a bit more care. 
In all twist sectors $\rho$ assumes a small constant,
bounded from below negative value 
$\rho \to - \kappa \simeq -10$
\cite{Barresi:2004qa,Barresi:2006gq}, therefore
indicating $\langle \mu \rangle \neq 0$ also for the full ergodic theory, i.e.
condensation of monopoles and confinement below $\beta^c_A$ 
\cite{Barresi:2004qa}. 
For every fixed $N_{\tau}~$ $\langle \mu \rangle_{erg}$ 
can be rescaled post-hoc to one through
$exp(\kappa \beta_A^c(N_{\tau}))$. This rescaling factor will
necessarily diverge for $SU(N)$, up to logarithmic corrections, 
like $N_{\tau}^{\epsilon \kappa}$ with $\epsilon= 2 \beta_0 (N^2-1)/N $, 
$\beta_0 = 11 N/(48 \pi^2)$ being the first coefficient of the 
$\beta$-function. For $SU(2)$ $\epsilon \kappa \simeq 1.4$.
This is of course not an obstacle in normalizing
$\langle \mu \rangle_{erg} = 1 $ in the
limit $N_\tau \to \infty$, although it remains a somewhat inelegant
feature of the Pisa disorder operator in the adjoint formulation. 
There is however a physical motivation for the non vanishing of $\rho$, 
as we will see in the following.

\subsection{Vortex free energy}
%--------------------------------
Having established the physical properties of phase II at finite temperature,
we will now turn to the 't~Hooft vortex free energy. As stated above,
this observable can only be calculated through a 
fully ergodic simulation.

\begin{figure}[thb]
\begin{center}
\includegraphics[angle=0,width=0.5\textwidth]{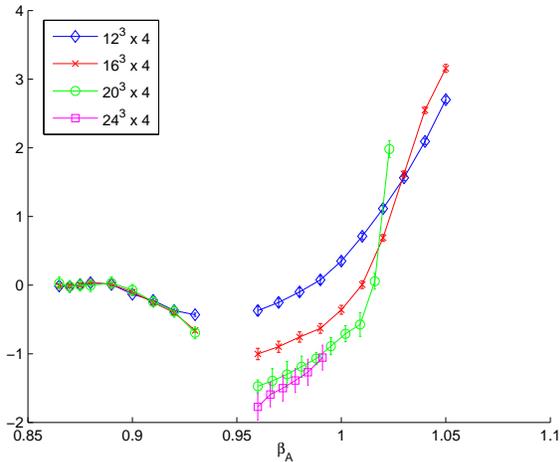}
\end{center}
\caption{Free energy 
$a N_\tau F_V$ along the $\lambda=0.8$ paths of Fig.~\ref{paths}.}
\label{free1}
\end{figure}

In Fig.~\ref{free1} the free energy of a vortex in lattice units is shown 
as a function of $\beta_A$ along the $\lambda=0.8$ paths of
Fig.~\ref{paths}.
The data points start right on top of the bulk transition and go on up to 
slightly above the finite temperature deconfinement transition. The 
behaviour up to the bulk transition is in agreement with the 't~Hooft vortex 
argument for confinement: if vortices behave "chaotically" ($F=0$) then 
the theory confines (phase I), while as deconfinement occurs 
$F\sim \tilde{\sigma} N_s^2$. As explained above, we cannot actually go too 
deeply into the deconfined phase, so we cannot
check if for $\beta_A \gg \beta_A^c$ 
the data are consistent with $O(N_s^2)$ plateaus or if they
saturate at some value in the thermodynamic limit, i.e. if a dual
string tension $\tilde{\sigma}$ can indeed be measured. 
To this purpose, assuming that the estimate in Eq.~(\ref{scale}) still
works at higher $\beta_A$, even taking into 
account that for higher volumes the asymptotic behaviour should kick in 
earlier, we would need to simulate around 50 parallel ensembles for each 
volume, again for a statistics of at least $O(10^5)$ per configuration in 
each ensemble. For volumes with $N_s \geq 20$, for which finite size effects 
start to be reasonably small, this goes beyond the computational power at 
our disposal, although it should be manageable with a medium sized PC 
cluster. A reliable estimate of $\tilde{\sigma}$ would be of extreme interest 
in light of the behaviour we find for $F$ in the confined phase 
of phase II, already reported in Ref.~\cite{Burgio:2006dc}. 
Vortex production is there clearly enhanced compared to 
phase I and the free energy stays negative up to the deconfinement
transition, where it rises to positive values. 

\begin{figure}[thb]
\begin{center}
\includegraphics[angle=0,width=0.5\textwidth]{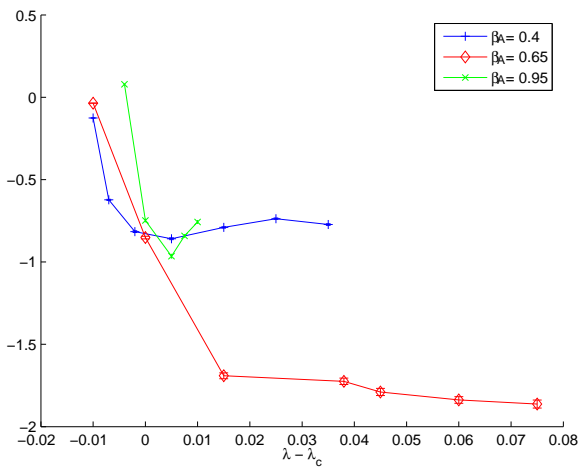}
\end{center}
\caption{Free energy $a N_\tau F_V$ for varying $\lambda-\lambda_c$ at fixed 
$\beta_A$ and lattice size $16^3 \times 4$ (see Table~\ref{tab6}).}
\label{low}
\end{figure}

Fig.~\ref{low} shows the free energy in a low $\beta_A$ confining region
well below the finite temperature transition. The negative 
plateau values away from the bulk transition at $\lambda_c(\beta_A)$
are consistent with what is observed in 
Fig.~\ref{free1} and with a vanishing free energy in the
limit $T\to 0$ \cite{Kovacs:2000sy}, since its value rises again for 
decreasing $\beta_A$ after reaching a minimum around $\beta_A = 0.65$. 
Larger volumes and a better extrapolation would be of
course needed to confirm this result.

As for the higher twists, although being proper only to the toroidal
topology they also show a surprising and interesting behaviour. Namely, we do 
not observe a free energy proportional to
the difference in topological index as one would have expected if the twist
observables $z_{i,4}$ were independent. We indeed observe a strong
correlation among the twist in the different planes, indicating a non trivial 
interaction among vortices; as a result in the confined phase, although the 
population of the $\pm1$ sectors for the single $z_{i,4}$ are comparable,
the distribution of the $\nu_k$ is such to follow the hierarchy 
$\nu_2/3 \gtrsim  \nu_3 \gtrsim \nu_1/3 \gg \nu_0$. E.g. for $N_s = 24$,
$\lambda = 0.8$ and $\beta_A = 0.96$ we find $\nu_0 = 0.020(1)$, 
$\nu_1 = 0.35(5)$, $ \nu_2=0.48(4) $ and $\nu_3 = 0.15(2)$. Such hierarchy 
is quite stable with the volume for $N_s \geq 16$. Taking into account that 
when calculating $F$ $\nu_1$ and $\nu_2$ need to 
be rescaled by a factor three to be compared with sectors $\nu_0$ and $\nu_3$, 
errors are still to high within the statistics at our disposal to allow a 
reliable measure of the free energy for the tunnelling other than from/to the 
0 sector. Approaching and crossing the deconfinement transition
the situation changes. The trivial sector starts
to dominate the partition function and the free energy to tunnel
from one sector to another becomes indeed proportional
to the difference in their topological index. Being the higher sectors however
exponentially suppressed in the deconfined phase their sampling requires
longer and longer runs as $\beta_A$ increases; the sampling
of sectors with topological index higher than one will become in practice
eventually unfeasible.

\subsection{$C^*$ boundary conditions, monopoles and $F$}
%--------------------------------------------------------

An interesting alternative check of our surprising negative value for $F$
is the evaluation of the vortex free energy for the modified action $S^M$ 
needed to define the Pisa disorder parameter in Sec.~\ref{mon_cond}. 
As already discussed in 
Sec.~\ref{sec2}, $C^*$ boundary conditions in the Euclidean time
direction $U(x+a N_{\tau}\hat{t}) = U^{*} (x)$ pose
no conceptual problem in our adjoint formulation. Any set
of twist matrices $\{\Omega\}$ in the fundamental representation
once projected onto the adjoint 
representation become gauge equivalent to periodic boundary conditions 
\cite{deForcrand:2002vs}. 
Given that any set $\{\Omega\}$ is gauge equivalent 
to the quaternion basis $\{\mathbb{I}_2, i \vec{\sigma}\}$ and since $C^*$ 
boundary conditions can be represented through the action of 
$\Omega_2 = i \sigma_2$, $U^{*}(x) = \Omega_2 U(x) \Omega^{\dagger}_2$, 
imposing them makes 
no difference in the dynamics of twist sectors: the configurations that
can be assigned to a given twist will simply be relabeled with respect 
to standard boundary conditions. In other words 
the corresponding combination of adjoint twist matrices, which would satisfy a 
given twist algebra when lifted to $SU(2)$, get reshuffled by the presence 
of $i \sigma_2$. A simple listing of combinations shows however that the 
number of states leading to the assignment of topological sectors 
$z=0,\ldots,3$ remains the same. 

$C^*$ boundary conditions alone therefore should not 
affect the value of $F$ for $S^M$. There
is however the bosonic field giving rise to the abelian monopole
through its non-trivial transformation property at the boundary
$\pi_2(SU(2)/U(1)) = \pi_1(U(1))=\mathbb{Z}$ 
\cite{DiGiacomo:1997sm,Frohlich:2000zp}; 
its r\^ole is far deeper and more interesting. For a Lie gauge group $G$ with
center ${{K}}$ and a maximal Cartan subalgebra ${{C}}$ 
it is well known that the abelian monopoles classified by 
$\pi_2(G/{{C}})=\pi_1({{C}})$ will carry center magnetic
charges classified by $\pi_1(G/{{K}})$ \cite{Lubkin:1963}; only 
the ones belonging to the kernel of $\pi_1({{C}}) \to\pi_1(G/{{K}})$
will be non singular \cite{Lubkin:1963,Coleman:1982cx}. For
$SU(N)$ this simply means that the $N-1$ abelian monopoles classified by
$\mathbb{Z}^{N-1}$ will also correspond to $N-1$ $\mathbb{Z}_{N}$ monopoles;
their corresponding Dirac strings (sheets in a 3+1 Euclidean formulation) will 
be in general {\it open} $\mathbb{Z}_{N}$ vortices.

In the simple case of $SU(2)$ the assignment is quite easy: to an abelian 
charge $n \in \mathbb{Z}$ will correspond a $\mathbb{Z}_2$ monopole of charge
${\rm mod}_2(n)\in \mathbb{Z}_2$. Only monopoles of charge $n= 2k$ carry no 
singularity along their world line; odd charged monopoles are singular in 
every gauge. The world line of the latter
saturates an open $\mathbb{Z}_2$ vortex, while only the former are
compatible with closed (or no) vortex sheets \cite{Frohlich:2000zp}. Our 
situation is however slightly different: $C^*$ boundary conditions take care of
magnetic charge conservation by making the unit charge abelian
monopole being put by hand into the system his own antimonopole; the Dirac 
sheet of the $\mathbb{Z}_2$ vortex starting on the monopole ends on itself. 
Such vortex is therefore 
spatially closed but cut through by a singularity along the world line of the 
abelian monopole. Moreover such ``cut'' cannot be gauged around;
the vortex is ``pinned'' down along one line by the singularity.
This should however not be detectable by $F$: the creation
of a further vortex will bring the system in a state which
for our purpose can aways be assimilated to either a 2- or 0-twist sector. 
This will make no difference below the bulk transition, where the partition 
function is dominated by open vortices anyway and all twist sectors
are equivalent: $F$ should still vanish there. In the deconfined phase 
$T> T_c$ the cost to create a further vortex should still scale like $N_s^2$, 
no matter what the background is, so that $F$ should become again large and
positive. Only in the confined phase of phase II, if the system prefers 
the vortex background, there should be a difference between $S$ and $S^M$:
if in the former $F$ is negative it should become positive in the latter.
Fig.~\ref{back} shows this to be the case:
\begin{figure}[thb]
\begin{center}
\includegraphics[angle=0,width=0.5\textwidth]{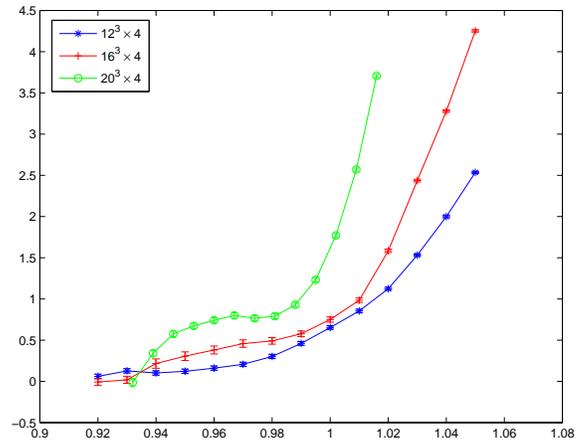}
\end{center}
\caption{$a N_\tau F_V$ for $S^M$ along the $\lambda=0.8$ paths of 
Fig.~\ref{paths}.}
\label{back}
\end{figure}
in the (singular) vortex background of $S^M$ the system 
prefers to make the creation of one further vortex
more expensive. Notice that the position of the bulk is slightly different
for $S$ and $S^M$, e.g. at $\lambda = 0.8$ it shifts from 
$\beta_A \simeq 0.89-0.90$ to $\beta_A \simeq 0.92-0.93$; starting our PT 
ensemble right on top of it we are able to span the range of couplings in 
one single run, reaching however the limits of our acceptance rate criterion 
for $N_s=20$. Why $F$ turns out roughly to be minus the half of what it
was for $S$ is not obvious. It might be related
to the non-standard character of the $S^M$ vortex background or to
the effective superposition of 0- and 2-twist sectors. In light of 
a recent proposal \cite{Rajantie:2005hi} such observable could also 
be linked to the direct evaluation of the monopole mass.

The previous discussion offers also an elegant explanation both to the 
sensitivity of $\rho$ to the bulk transition and to its non-vanishing 
in the low $\beta_A$ limit of phase II observed in 
\cite{Barresi:2004qa,Barresi:2006gq} and already discussed in 
Sec.~\ref{mon_cond}. Action $S$, consistently with what is expected in the
continuum limit, completely suppresses the presence of $\mathbb{Z}_2$ 
monopoles,
allowing only topological, i.e. closed, vortices. The construction
of $\rho$ we have adopted, following \cite{DiGiacomo:1997sm,Barresi:2004qa}, 
uses a unit charged abelian monopole, which as discussed above introduces a 
singularity exactly 
through the presence of a $\mathbb{Z}_2$ monopole. The physical states
described by $S$ can therefore never be equivalent to those of $S^M$ and 
hence the discrepancy. This was
already ``predicted'' in Ref.~\cite{Frohlich:2000zp}, where an alternative
construction using {\it even} charged monopoles was proposed as the only
one fully consistent with the continuum pure Yang-Mills action.
Singular gauge configurations will not be allowed there and only genuinely
closed vortices will exist. Such modified construction 
should therefore be the obvious choice for adjoint actions and its 
feasibility might be worth to explore.
The discrepancy in $\rho$ observed here cannot be detected through
actions transforming with the fundamental representation: although they 
should disappear in the continuum limit, $\mathbb{Z}_2$ monopoles are still 
present both with and without monopole background in the range of 
parameters commonly used in simulations, while the topological structure 
of vortices is anyway blurred by the fixed boundary conditions.

\section{Electric flux and vanishing of $F$}
%--------------------------------------------

A few comments are in order to clarify some properties of $F$ 
discussed in the literature which could 
seem to be in conflict with our result. 
In Ref.~\cite{'tHooft:1979uj} an exact duality relation 
between electric and magnetic fluxes implying the vanishing of $F$
was proved under two main assumptions:
first one must be able to define a set of regularized operators in
the fundamental representation; second the limit $T\to 0$ has to be taken 
after the finite temperature
compactification of the time direction. As for the latter, we actually
agree with $F = -T \Delta S \to 0$ as $T \to 0$, which 
is almost obvious in our formulation.
The former condition has more far reaching consequences. It was extended 
to finite temperature and is indeed the key point 
of the derivations of many interesting dynamical relations in 
Yang-Mills theories, mostly
using reflection positivity for the fundamental Wilson action on the
lattice \cite{Tomboulis:1985}. In an adjoint discretization however
observables like the fundamental Polyakov loop, the fundamental trace 
of the electric flux free energy etc. needed to prove such relations
are all undefined. Formally their expectation
values and all their correlators vanish identically.

This can be better understood within the Hamiltonian formulation of lattice
Yang-Mills theories, adapting the exact construction of the 
$SU(2)$ Hilbert space given in Ref.~\cite{Burgio:1999tg}. Since the 
irreducible 
representations of $SO(3)$ are simply the integer representations of $SU(2)$,
the Hilbert space of $SO(3)$-invariant states in each twist sector is 
given by the subset of the $SU(2)$ states described in 
Ref.~\cite{Burgio:1999tg} 
(with corresponding t.b.c.) with 
{\it all} links and intertwiners labelled by integer spins {\it only}. 
Since topological sectors are automatically accounted for in $SO(3)$ 
with p.b.c. we have 
$\mathcal{H}_{SO(3)} = \bigoplus_i \mathcal{H}^{i}_{SO(3)}$, $~i~$ being the 
winding number corresponding to the twist sectors, with 
$\mathcal{H}^{i}_{SO(3)} \subset \mathcal{H}^{i}_{SU(2)}$ the fixed twist 
spaces mentioned above. It is now straightforward to see how
fundamental operators annihilate the Hilbert space of $SO(3)$-invariant
states.
Let us illustrate this by an example. Take a fundamental Wilson loop,
winding or not around the boundaries: on the 
$SU(2)$ Hilbert space such operator 
can be represented as a closed string of spin 1/2 located on the corresponding 
links; as proven in Ref.~\cite{Burgio:1999tg} its action on 
a generic $SU(2)$ state will generate states where the representations 
labelling the links and intertwiners of interest are simply composed
with the 1/2 representation via the common spin composition rules. Since
a state in the $SO(3)$-invariant subspace defined above will only carry 
integer spins, its composition with our operator will generate states
which will necessarily carry semi-integer representations on the links
of interest, i.e. it will not belong to $\mathcal{H}_{SO(3)}$ anymore.
Generalizing, the action of a fundamental operator $\mathcal{F}$ on a state 
$\psi = \sum_i \lambda_i \psi_i$, $~\psi_i \in  \mathcal{H}^{i}_{SO(3)}$ 
will generate states living in the orthogonal 
complements of $\mathcal{H}^{i}_{SO(3)}$ in $\mathcal{H}^{i}_{SU(2)}$, 
$\mathcal{F} \psi_i \in \mathcal{H}^{i, \bot}_{SO(3)}$ with 
\begin{equation}
\mathcal{H}^{i}_{SU(2)} =  \mathcal{H}^{i}_{SO(3)} \bigoplus 
\mathcal{H}^{i, \bot}_{SO(3)}\,.
\label{hilbert}
\end{equation}
When the dynamics can be described by a Hamiltonian 
$H =E_c+V$ transforming under the adjoint representation\footnote{Only 
the transformation properties of $V$ need to be specified; $E_c$ is always 
diagonal with our choice of basis \cite{Burgio:1999tg}.}, as it is 
the case for Eq.~(\ref{ouraction}), it is legitimate to 
restrict the whole treatment to $\mathcal{H}_{SO(3)}$. Then obviously
$\mathcal{F} \mathcal{H}_{SO(3)} \equiv 0$, which is by definition 
the only state in the intersection 
$~\mathcal{H}^{i}_{SO(3)} \bigcap \mathcal{H}^{i, \bot}_{SO(3)}$. 

A trivial consequence of the above arguments is that with an 
adjoint action reflection positivity constraints can only 
be invoked for adjoint observables, ensuring e.g. that adjoint Polyakov loop 
correlators will be positive definite;
constraints derived from fundamental operators will be invalid. 
This concerns e.g. the reflection posi\-ti\-vi\-ty constraints
among fundamental Wilson loops, static quark potential, electric flux free
energy and vortex free energy (see e.g. Appendix I in 
Ref.~\cite{Tomboulis:1985}).
In particular this latter constraint is interesting since it 
could seem to 
contradict our result for $F$. Such relation between the Fourier transform 
of the vortex 
free energy and the electric flux free energy, essential to derive the 
vanishing of $F$ in the confined phase, is ill defined in an adjoint 
theory, since it needs the action of a fundamental maximal Wilson loop winding 
around the space boundary to be established, as in Eq.~(4.6) of  
\cite{'tHooft:1979uj}.
With an adjoint weight in the partition function Eqs. (4.9) and (4.10) of the 
above reference become identically zero.
Therefore the operator there given in Eq.~(5.2), although still well defined, 
cannot be related to the projector onto a state of fixed electric flux.
Also the alternative definition given in Ref.~\cite{deForcrand:2001nd} through 
fundamental Polyakov loops $~L_F~$ modified via a twist eater at the time 
boundary
is only valid in a fundamental theory (with twisted boundary conditions). In
an adjoint theory the first line of their Eq.~(15) cannot be inverted.
Alternatively, correlators of $L_F$ will in general vanish identically giving
no useful bound on their sign via reflection positivity for the only
potentially non-vanishing case of maximal displacement. There is
therefore no guarantee that the right hand side of their Eq.~(16) will be
positive, i.e. that it can indeed be interpreted as the exponential of a free
energy. 

An electric flux operator can of course be defined also in our adjoint model:
it will simply be given by an {\it adjoint} maximal Wilson loop winding 
around the space boundary. Such operator cannot however be related to the 
vortex free energy as defined in Eq.~(\ref{free_en}).

\section{Conclusions}
%--------------------
In this paper we have studied on the lattice at finite temperature 
a pure $SO(3)$ gauge theory, which transforms 
under the actual gauge symmetry group of pure $SU(2)$ Yang-Mills 
in the continuum. Extending the analysis in Ref.~\cite{Burgio:2006dc} we have
employed the Pisa disorder operator for monopole condensation to 
establish the properties of the theory within
the weak coupling phase II, which allows a well defined continuum limit,
finding confined and deconfined phases 
separated by a transition of presumably second order
consistent with the universality class of Ising 3-d.

The vortex free energy $F$ is however found not to vanish at $T \neq 0$
in the confined phase of $SO(3)$. As discussed above, standard arguments 
for its vanishing up to $T_c$ cannot be applied when working with an 
adjoint action, so that our result is in itself not contradictory. 
Moreover, the vanishing of $~F~$ in the confined phase is in general
just a sufficient condition for confinement. The
full adjoint theory discussed here possesses neither
center symmetry nor well defined fundamental observables. Only
adjoint observables make any sense, a fundamental string tension being 
impossible to calculate. 
There seems therefore no compelling physical reason for the 
't~Hooft vortex free energy to vanish for $T\neq 0$, since this would 
be linked to an area law behaviour of 
fundamental Wilson loops, i.e. to the existence of a fundamental string 
tension, which however loses its meaning as soon as the specific properties of 
semi-integer discretizations are lifted. Indeed, it would be surprising to 
establish the existence of an order parameter for the breaking of a 
symmetry the theory does not possess \cite{Holland:2003jy}. This however 
does not in our opinion violate universality, since such properties are
not essential to describe the dynamics of continuum {\it pure} 
Yang-Mills theories.
Only physical properties like deconfinement
temperature and universality class or the glueball spectrum are 
preserved independently of the discretization used. Only those should
therefore be addressed in trying to establish a topological mechanism of 
confinement valid both in pure Yang-Mills theories and in full QCD. 
If no symmetry breaking and therefore no order parameter 
is available the properties of the phase transition can still be established
through the specific heat or other thermodynamic observables  
\cite{Datta:1999di,Datta:1999np}.

The non-vanishing of $F$ implies that the dual string tension cannot 
serve as an order parameter for the adjoint theory. For this purpose
it would need to vanish exponentially in the thermodynamic limit in the 
confined phase, while our simulations indicate that it will vanish at most 
as $O(N_s^{-2})$. 
Our results show therefore that between 
the monopole condensation parameter and 
the 't~Hooft vortex free energy (i.e. a dual string tension) only the former 
seems to retain all its order-parameter character independently of the 
discretization used, at least once rescaled to assume a constant value in the
confined phase. This is a simple consequence of its property
to vanish exponentially {\it above} the critical temperature, 
as e.g. the magnetization for a ferromagnet. The rescaling should anyway
become unnecessary if one adopts the alternative prescriptions given in
\cite{Frohlich:2000zp}.
Monopole condensation can therefore play the r\^ole of an order parameter
both for the $SU(2)$ invariant quenched theory, possessing
center symmetry but no vortex topological sectors, and
for the $SO(3)$ invariant pure Yang-Mills theory, where center symmetry is 
absent. Vortex topology does not provide there a suitable order parameter.

Why the 0-twist sector gets strongly suppressed in the
confined phase causing vortex free energy to take a negative value, 
i.e. why the Yang-Mills
action finds it energetically favorable to create at least one 
vortex, and why twist observables in different directions hint at a non
trivial interaction pattern among vortices are questions which need to be 
answered and that we will try to address in the near future.

\section*{Acknowledgements}
%-------------------------
We thank A. Barresi, M. D'Elia, P. de Forcrand, T. Kovacs, M. Pepe and 
U.J. Wiese for comments and discussions. G.B. wishes to acknowledge 
support from INFN.

\bibliographystyle{apsrev}
\bibliography{bib.bib}

\end{document}